# Talking about interaction*

Stuart Reeves, School of Computer Science, University of Nottingham

Jordan Beck, College of Information Sciences and Technology, Penn State University

**Abstract**

Recent research has exposed disagreements over the nature and usefulness of what may (or may not) be Human-Computer Interaction's fundamental phenomenon: 'interaction'. For some, HCI's theorising about interaction has been deficient, impacting its capacity to inform decisions in design, suggesting the need either to perform first-principles definition work or broader administrative clarification and formalisation of the multitude of formulations of the concepts of interaction and their particular uses. For others, there remain open questions over the continued relevance of certain 'versions' of interaction as a useful concept in HCI at all. We pursue a different perspective in this paper, reviewing how HCI treats interaction through examining its 'conceptual pragmatics' within HCI's discourse. We argue that articulations of the concepts of interaction can be a site of productive conflict for HCI that for many reasons may resist attempts of formalisation as well as attempts to dispense with them. The main contribution of this paper is in specifying how we might go about *talking of* interaction and the value of interaction language as *promiscuous concepts*.

# 1. Introduction

Concepts of 'interaction' have been used in HCI as ways of talking about the myriad forms of relations that emerge amongst and between people living with and around digital devices and systems. Sometimes, the wide range of senses in which concepts of interaction are used in HCI is a source of confusion. There are many different and potentially incommensurate ways of constituting and schematising what we mean when we say 'interaction', 'interacting', 'interactivity', and so on. Looking to this journal alone, for example, yields the following: multimodal interaction (Deng et al., 2017), sonic interaction (Rocchesso & Serafin, 2009), ubiquitous interaction (Kostakos & Musolesi, 2013), touch interaction (Ghosh et al., 2019), and interaction with virtual environments (Wilson, 2005). For some in HCI, this has become problematic.

To be clear, we will *not* address that issue in this paper. Instead we chart a different course by unpacking what the concepts of interaction let HCI *do*. One's understanding of a concept constrains and influences the way that concept can be worked with (Beck & Ekbia, 2018). We do not think that this has yet been articulated in response to HCI's "*current turbulence*" (Kuutti & Bannon, 2014).

In order to head off confusions, we are going to asterisk 'interaction'[1] to remind the reader (and ourselves) that we are interested *not* in addressing what 'interaction'—its various versions included—might be in and of itself. Rather, interaction* is about bracketing the matter, i.e., considering, in a broad sense, HCI's *concept(s) of interaction and the conversations about them*. So, this is what we mean when we say interaction* here: so instead read 'the concepts of interaction'.

Recently, researchers have been taking stock of the import of interaction* for HCI's disciplinary development. These positions can be construed as divergent. For some there is a case being made for the continued relevance of interaction* to HCI through better formal definition. For others, there is frustration about the limitations of HCI's renderings of interaction* and an implied dismissal of interaction* from the realm of the cutting edge. The broader tone of these debates thus can be seen to split in two potentially divergent directions. The former rests on the implication that, although HCI's research uses interaction*'s language and its variants frequently, we do not really know what it is that we mean, and thus there is a need for definitional work. The latter makes a case for new conceptual apparatus entirely.

---

[1] To clarify, we are not addressing any versions of interaction directly, nor do we intend a specific sense of or perspective on what *we* mean by 'interaction'. This, we feel, has been done far more comprehensively than we ever could in this article, and so we refer the reader to treatments by Hornbæk and Oulasvirta (2017), Janlert and Stolterman (2017a, 2017b), Dubberly et al. (2009), etc. We hope that our bracketing of interaction foregrounds some of the reasoning HCI researchers are employing so as to produce an objective stability to the concept in use (Garfinkel, 2002, pp. 30-33).

Yet, this debate raises questions which are perhaps even more fundamental: What are recent discussions about interaction* doing for HCI research communities? Where might they spring from? What could we mean when we talk about interaction* in HCI? How does HCI talk about interaction* and its associated or implied language families? How might we evaluate proposals to define or dismiss interaction*? Are there other ways we can think about interaction* and its role in HCI? In short, we need to talk about how HCI approaches debates on interaction*[2], just as much as some might feel developing particular concepts of interaction as a viable topic for HCI's work.

We offer an argument that proposes interaction* and its languages as *promiscuous* concepts that—in spite of potential negative connotations of this term—denotes interaction*'s positive, actively generative, wholesale permissive character. Interaction* acts as a pragmatic—but not infallible or unbreakable—'gravitational force' that holds together the shifting sands of HCI's research communities just enough so as to be useful. In being this way, interaction* offers opportunities for new associations or affiliations. In doing this we want to foreground interaction* (the idea, its discourse) to surface what we feel are various submerged misunderstandings and crossed wires.

We begin by describing some motivating issues. These suggest why, we think, discussions of interaction* have taken hold in HCI in recent years. We then outline the broad shape of emerging interaction* debates that suggest different disciplinary pulls. Finally, we make some arguments for retaining interaction* by reframing its troublesome nature as a matter of language. Finally we sketch ways for HCI to cope with this through a therapeutic approach informed by understandings of ordinary language.

---

[2] To translate the bracketing again for the utmost clarity: i.e., we call for HCI's reflection on how *its discourse* on the concepts of interaction plays out.

## 2. The problem with interaction*

Recently in HCI, interaction* has (re)surfaced as a focal topic in its own right. Communities of HCI researchers' treatments of interaction* are multitudinous, however we restrict our view here to two *seemingly* oppositional treatments (the situation is more complex, as we argue later, as they represent an evolving set of conversations rather than coagulated positions). To summarise we detect two broad directions. 1. 'Defining': seeking to formally arrange descriptions of a specific perspective on interaction* or schematise perspectives on interaction* with a view to establishing some measure of agreement in HCI (including its boundaries). And 2. 'Dispensing': a tendency to argue that clear conceptual constraints within HCI's interaction* discourse suggests the abandonment of interaction in some fashion or another. (This is *our* particular reading of the debate, and the reader of this paper should bear that in mind.)

### 2.1 Defining interaction*

As Hornbæk and Oulasvirta—hereon, H&O—argue in their 2017 CHI paper, while "the term interaction" is often held to be "field-defining" and a "workhorse", it is, in their view, "underdefined". Accordingly, H&O's paper is perhaps the most comprehensive of recent work attempting to address interaction* in HCI. In doing so it offers something of a catalogue approach to what H&O see as a problem of profound historic disinterest in systematic articulations of interaction* in HCI. By catalogue approach we mean that H&O advance their argument by presenting a range of different "concepts of interaction" (i.e., interaction*) that they contend are largely inexplicit, submerged or "underdefined" (in H&O's words), but nevertheless drive different threads of HCI's research. These are set in contrast with the "folk notions of interaction" (ibid., p. 1) as presented in HCI textbooks wherein H&O argue that one might reasonably expect a "definition or high-level discussion of interaction" (ibid., p. 1), and yet nevertheless fail to locate such a thing[3]. For example, we might consider how particular senses of interaction* can be a way of speaking in a language that reasons about models of stimulation and response between people, while in another mode of working interaction* be conceptualised as an interpretive process performed by members of social groupings (in accord with particular psychological and sociological influences in HCI respectively). There are many more examples like these, which are documented by H&O. Overall the schema H&O present seems to stand as a reaction to the many ways of talking about interaction* which can be 'at play' at any time in no distinctly differentiated way within HCI's discourse, yet with potentially highly substantive differences in use and sense for the communities which employ them.

---

[3] While believe this pursuit to be the result of a confusion, we temper this with the point that there *are* times when looking for more definition can be generative.

The purpose for HCI, state H&O, is thus to "move from a folk notion of interaction to a notion that really *explains* interaction" (ibid., p. 7). This means locating *causal* reasoning that accounts for matters such as "how intentions are formed or affected by interaction" (ibid., p. 8). Overall, the absence of definition work in HCI seems to be a key motivating factor for H&O, with the view that practicing such definition work and developing fully causal explanations for interaction* will lead to an increase in HCI's "problem-solving capacity" (Oulasvirta & Hornbæk, 2016).

Some of these concerns also emerge in contemporary work. Janlert and Stolterman (2017a)[4]—hereon, J&S—mirroring H&O's complaint about "folk notions of interaction"—suggest that HCI's understandings of interaction* rely too much on "common intuitions"[5] and that "interactivity"—J&S's focal interaction* word—needs to be approached "in a systematic and analytical fashion". Although J&S hang their address of interaction* upon the question of "increasing interactivity"—that a key feature of designed systems is in providing different amounts of "interactivity"—ultimately, they return to the motivations of H&O by building up definitions and terms, albeit in a way that does not clearly consist of a plurality of alternatives, unlike H&O's catalogue. In many ways, J&S present a particular but very explicitly stated attempt at a figuration of some sense of interaction* that could conceptually reside somewhere within H&O's schema.

---

[4] We also refer the reader to Janlert and Stolterman's book *Things That Keep Us Busy: The Elements of Interaction* (Janlert & Stolterman, 2017b) which also seeks further definition of interaction*.

[5] This is similar to what Galle would call intuitive resonance (Galle, 2011, p. 82). In brief, people think that the term will be understood by their audience and so they do not offer a definition in an effort to preserve resources (e.g., page space) for other purposes.

As noted by H&O, explicit attempts at core definitional work remain sparse. One exception we would point to is a cybernetic conception articulated by Dubberly et al. (2009). This is more consonant with the interaction design orientation of J&S[6] yet probably falls within the "interaction as control" conceptual orientation covered by H&O (even though Dubberly et al.'s paper is not cited by H&O). Dubberly et al. directly approach what they mean by "interaction", arguing that, in an interaction design sense, their way of articulating "interaction" is as a "way of framing the relationship between people and objects designed for them—and thus a way of framing the activity of design" (p. 96). Like H&O and J&S, then, Dubberly et al. seek to firm up a particular sense of interaction* through greater conceptual definition.

Softer still than definitional work is the staking out of specific communities. This includes attempts to foreground something like agreement for interaction* that sets of HCI scholars approach in a consistent way (i.e., either recognising or developing a common orientation towards what constitute relevant phenomena for a community / perspective, such as sonic interaction (Rocchesso & Serafin, 2009) or ubiquitous interaction (Kostakos & Musolesi, 2013)). Perhaps the clearest programmatic statement of this sort may be found in recent arguments for, or at least recognition of an extant "interaction science" or "science of interaction"[7] in HCI. This has been pursued both by a Special Interest Group meeting (Howes et al., 2014) and "Spotlight" at CHI 2014[8], undergirded by the parallel commencement of a Journal of Interaction Science in 2013 (Bahr, 2013).

**2.2 Dispensing with interaction*?**

For us, H&O's paper in particular sits quite clearly in response to prior work by Taylor ("After Interaction" (Taylor, 2015)), Verbeek ("Beyond Interaction" (Verbeek, 2015)), and Kuutti and Bannon ("The Turn to Practice in HCI" (Kuutti & Bannon, 2014))—hereon K&B.

---

[6] It seems common to conflate (or at least not clearly distinguish) interaction design and human-computer interaction in discussions of interaction*; e.g., see Svanæs' (2013) chapter "Philosophy of Interaction" in https://www.interaction-design.org/literature/book/the-encyclopedia-of-human-computer-interaction-2nd-ed/philosophy-of-interaction

[7] We note the distinction here for these are potentially two different programmes.

[8] See https://chi2014.acm.org/spotlights/interaction-science for more details (verified May 2019).

Collectively, they offer serious challenges to a definition-driven model of interaction*. Taylor and Verbeek put forward a set of convincing arguments about the possible ways in which interaction* might limit, and in some cases repress, HCI's potential and scope. K&B, on the other hand, argue more for the recognition of a pre-existing "Practice paradigm" that can complement what they label as the existing "Interaction paradigm" in HCI.

Taylor argues that expressions of interaction* in HCI have been tied conceptually to the materiality of user interfaces as well as conceptual configurations of human-machine binaries. He states that interaction*'s formulation in this way has led HCI to "concentrate [...] attentions on the interface" to the exclusion of other things that HCI correspondingly lacks the conceptual apparatus to deal with, such as broader structural-societal issues. Verbeek argues that the concept of "mediation" may be more apposite for design, while Taylor suggests that design constructs complex entanglements between human and machine to constitute "*worlds*" rather than merely "discrete interaction[s]"[9] of mainstream HCI. What kind of "worlds" HCI could or should be building is the question: current interaction* in HCI hinders deeper thinking on that matter. Taylor and Verbeek thus find HCI's senses of interaction* to be insufficient and constraining. In many ways we concur.

One way of misreading the 'interaction* is deficient' view is by characterising it as an argument *against* what Taylor posits as the prevalence of Engelbartian interaction—a reference to Engelbart's famous "Mother of all Demos" which Taylor argues set the scene for HCI's 'default' sense of interaction*. Engelbartian interaction, suggests Taylor, "prefigures an interface, foregrounding a very particular set of relations between user and computer" that is naturally limited in its scope. H&O appear to take Taylor in the mode of rejecting Engelbartian interaction; suggesting that it must be done away with in order to *correct* HCI's design role as "mediator" and focus on the enactment of "worlds" through design.

However, we take Taylor's argument to be more complex than that. Taylor argues that Engelbartian interaction is "taken for granted" in HCI and interaction design, but then appears to generalise this by noting that "as a concept, *interaction* hinges on an outmoded notion of technology in use" (emphasis added). On the one hand, this could be read as a challenge to an imagined status quo—a unitary sense of specific Engelbartian interaction—and, on the other, "outmoded notion" might apply to whatever *other* senses of interaction* we have in mind, which, as H&O point out, is a broad set of possibilities. Hence, there is potential for a pejorative characterisation of HCI's senses of interaction* and, indeed, this is moved upon by H&O.

---

[9] The discussion ensuing from Taylor's article has been documented online (https://ast.io/back-to-interaction/), which itself initiated a subsequent workshop at Microsoft Research Cambridge in 2016 (see https://ast.io/promiscuity-of-interaction/ for a report).

Verbeek's suggestion to go "Beyond Interaction" is similarly vexed. The provocative title is not matched by provocative content in the article. Verbeek is actually somewhat muted in that he argues both that "interaction might not always be the most helpful concept" for design *and* that, considering present deficiencies in interaction*, newer concepts of "technological mediation" are "shedding new light on the field of interaction design". This feels more pluralistic than the way H&O seem to take it. However, we can see why "Beyond Interaction" has the potential to be treated as a way of dispensing with interaction* entirely.

K&B on the other hand present the "turn to practice" within HCI as the gradual establishment of an alternate to the "Interaction paradigm" which has "tended to focus on momentary and ahistorical HCI situations". While K&B underline the importance of the "Interaction paradigm" that is "brimming with unresolved problems", and therefore distance themselves from a dispensing view, they do present an argument with a shared genetics to Taylor's and Verbeek's. K&B point to the limitations of the "Interaction paradigm" and the need for a "decentering" of the "privileged position of interaction", resulting in a focus on practices instead:

> For the Practice paradigm, a whole practice is the unit of intervention; not only technology, but everything related and interwoven in the performance is under scrutiny and potentially changeable, depending on the goals of the intervention. Thus the changing technology is but one of the options. (Kuutti & Bannon, 2014)

While Taylor, Verbeek, and K&B's interventions feel productive for HCI, we are cautious about readings that adopt even a mild dispensing view of any sense of interaction*. Even K&B's respectful account of two "paradigms" in HCI tends to put them into a relationship that might be construed as antagonistic. Furthermore, while Engelbartian interaction may have "become a cornerstone for HCI and interaction design (IxD)" as Taylor writes, and though it seems reasonable to concur with Verbeek that developing new notions such as "technological mediation" seem fruitful, H&O's response puts forward a comprehensive case that these are potentially impoverished and somewhat, probably unintentional, straw man views of the sheer diversity of interaction* present in HCI. Yet, as we have made clear, this is nevertheless done by H&O under the auspices of seeking further definitional and schematised clarity about interaction*, about which we exercise caution.

To some extent, Verbeek's argument also tends to be parseable as constructing a straw man of interaction* in HCI by not giving a clear sense of what is meant when he uses "interaction". The implication is probably a specific "interaction as dialogue" sense, but the reader is left wondering, unlike Taylor's piece where the argument singles out HCI's overreliance on defaulting to Engelbartian interaction which is fortunate enough to have a clear cultural reference for its own definition. Due to ambiguity, Verbeek's characterisations of interaction may then be criticised as insufficiently expressive. However, like J&S, once again the language of implicitness, agency, control, etc. all surface here as underlying driving concerns. (So, there *are* points of common ground in our sketch of the different ways of treating interaction*.)

This also points to an issue with our initial binary categorisation, between what we have called the 'interaction* defining' and the 'interaction* dispensing' tendencies. We are *not* arguing that H&O, J&S, Taylor, Verbeek, or K&B are definitively motivated by this, intended the effect, or would necessarily agree with our characterisations. K&B, for instance, explicitly state they are not suggesting what they call the "Interaction paradigm" should "disappear". Rather, we detect implications about where the state of HCI's interaction* discussion could lead, as perhaps any reader of their work *might*. And this is worth exploring and road-testing if they are to be adopted by other HCI researchers in the future.

In a sense HCI's discourse on interaction* is about authors' divergent reactions to 'discipline': either as agents of 'disciplining' or instead feeling the effects of *being* disciplined. In other words, rejecting the usefulness of interaction* in whatever form could be seen as a response to the constraints felt from the kinds of disciplinary articulations of HCI we have discussed in this paper (e.g., that HCI has concepts X, Y, Z and A, B, C ways of applying them, thus constituting HCI's agreements on interaction*). When we foreground this largely unstated dynamic—restating the debate on interaction* as an *effect* of disciplinarity itself—we think it sheds light on the frictions at play in a way that we imagine few would want to explicitly sign up to as the intended effect.

We have tried to illustrate how the current discussions at times appear to suffer from cross-purposes and ambiguity about what might even be meant by interaction*. We also suggest that this discussion lacks a clear articulation of the position we present here: interaction* as a kind of generative broker between a complex mix of interlocking communities that form HCI research.

We will next try to excavate the disciplinary milieu in the midst of which this discussion takes place. By unearthing its features, we can explore the ways it lends a framing to H&O, J&S, Taylor, Verbeek, and K&B's treatments of interaction*. In doing so we will attempt to move further still from characterising this discourse as between two opposing solutions and rather as more broadly symptomatic of the state of affairs in HCI research.

## 3. Why interaction* now?

So why this increased focus on interaction* now? We can frame a possible answer by elaborating a cluster of interrelated concerns floating around HCI presently and historically. These include: multiplying sub/supra-communities, increasing intellectual differences, shifting senses of purpose, negotiating academia and industry, questioning disciplinarity, and doing away with old intellectual concepts and adopting new ones. Many of these are not exclusive to HCI but they are likely exacerbated by the peculiar makeup of its overlapping communities and particular confluence of inputs (cf. Grudin, 2017). We also admit we likely have an ACM SIGCHI bias.

*Multiplying Sub/Supra-Communities.* In recent years HCI research communities have been changing along various dimensions. Taking the ACM CHI conference as a bellwether[10], there has been significant community growth in terms of sheer *participation* rates rather than attendance. Materials submitted for CHI in the early 2000s hovered at ~400-500 items, whereas CHI 2018 has seen a rise to over 2500 items[11], in spite of present attendance (2010s onwards) being similar to or only slightly higher than those seen during the dotcom boom of the late 1990s. This disparity of growth between participation and attendance plays into community divergence as new areas and interests come online to service smaller sub/supra-communities, which are then accommodated by a growing number of conference subcommittees. We speculate that increased participation could play into greater visibility of divergent languages of interaction* (which perhaps precipitates H&O's paper).

---

[10] We fully accept that the use of the CHI conference in this way is a limitation to this point in the argument given that it does not take into account other HCI communities of research.

[11] See https://sigchi.org/conferences/conference-history/CHI/ (verified May 2019).

*Increasing Intellectual Differences.* HCI's research communities have become increasingly intellectually diverse over the past decade, including a range of new initiatives such as fabrication and making, postcolonial computing, labour rights, inclusion and diversity, AI and HCI, critical and speculative design, and digital civics, among others. As Rogers puts it, this has been a process of HCI "recasting its net ever wider" (Rogers 2012, p. 3). Such recasting has introduced new domains, approaches, agenda, research styles, and different contribution types. And with this comes an increase in the range of possible points of difference between members of HCI communities, and correspondingly alternatives (and perhaps, challenges) to normative, unstated stances on interaction*.

*Shifting Sense of Shared Purpose.* Perhaps in response to these developments, concerns are raised by some HCI researchers about a lack (or loss, depending upon one's point of view) of coherent shared purpose in HCI research (Beck & Stolterman, 2017). For example, Liu et al. (2014) and Kostakos (2015) make the case for HCI having an absence of what they call "motor themes"—jointly oriented-to endeavours within HCI that establish HCI as a "normal science" (Kuhn, 1970)[12]. The perceived absence of coherence reflects discontiguities in researchers' orientations about what kind of thing HCI is as a disciplinary object (Blackwell, 2015). In other words, coherence and shared purpose debates seek to resolve broad, perceived conceptual problems for HCI, of which we argue interaction* is but the latest iteration of. H&O make just this connection in their 2017 CHI paper, identifying Liu et al.'s work as a motivating factor in their topicalisation of interaction*.

*Academia and Industry.* Related to the shifting senses of HCI's purpose is a complex set of challenges based around the *political* relations of HCI's various community members and their research agenda. Initially narrowly focussed on improving usability, expansion towards various "grand challenges" (Shneiderman et al., 2016) such as development goals has ensued. It is not unusual to hear members of HCI communities expressing desires to positively effect change in the world through HCI's design practices. Widespread adoption and the structural-societal (economic, social, political) embedding of HCI-relevant technologies has continued apace, and yet although such technologies touch almost every part of everyday life, their design is dictated primarily by large technology corporations and (vigorously invested) startups, as opposed to alternate models, e.g., government or hybrid partnerships as in Scandinavian participatory design traditions. Most notably, social media (but also other network technologies) has played a significant role in this, backed by the massive accrual of personal data to private entities, and most recently, increased application of Machine Learning based systems to classify and sort the human subjects of such systems (e.g., face recognition). Yet the corporations involved deeply in driving many of these changes *are* represented in HCI communities (often blandly referred to with the gloss 'industry'). This creates further frictions in the configurations of relationships between different HCI community members' orientations as to what constitutes 'good' interaction\*.

---

[12] The Interaction Science Spotlight (Howes et al., 2014) provides an interesting contrast to this call for coherence; for "Interaction Science", it is researchers' *existing* sense of alignment and agreement that is being brought to prominence *rather* than the perceived need for alignment and agreement itself doing the driving (as in the case of (Liu et al., 2014; Kostakos, 2015)).

*Questioning Disciplinarity.* As digital technologies—products and services and systems—have become widespread, so naturally the interests of *other* fields and disciplines towards investigating HCI-relevant matters has increased. Calls for interdisciplinarity frequently drive research funding structures. Conceptualising HCI as a discipline (as has been argued against) can lead to a sense of stakes (or perhaps rights) to some particular intellectual territory by HCI communities which is increasingly 'trespassed' by 'others'. This point is as much about how HCI draws its own boundaries (interaction\* being a case in point) as how communities operating within them in some capacity notice such matters.

*Doing Away with Older Concepts.* We have a collection of historical accounts of progress and change in HCI research. Such accounts might articulate this history as forming "waves" (Bødker, 2006), "turns" (Rogers, 2012) or "paradigms" (Harrison et al., 2007). Of these, it is perhaps primarily the use of "paradigm" with respect to HCI—with its allusions to Kuhn's well-known description of scientific revolutions (Kuhn, 1970)—that might *suggest* most clearly a connection with ideas of doing away with or superseding HCI concepts that are no longer of use, such as interaction\*. However, this is not what seems to have been meant by Harrison et al. in that they argue "shifts in the underlying metaphor of interaction" for HCI means that "new paradigms do not disprove the old paradigms, but instead provide alternative ways of thinking" (p. 3). Nevertheless, once used, the Kuhnian sense of "paradigm" tends to prevail in spite of the authors' appeals within the paper (a strategic error, perhaps, in including such a term within the paper's title).

**12**

In sum, this cluster represents significant concerns: about the model(s) HCI researchers express a desire to operate along the lines of, about what HCI's shared values are along which researchers jointly travel, about how members of HCI communities relate to one another and upon what basis those relations are conducted, about how HCI 'stays together' as sets of communities with conflicting pulls over interaction*, and, finally, about how HCI might even tell a 'coherent' and agreed-upon story of its history when there are multiple ways of squaring what the role of interaction* is in that.

If HCI communities cannot tell more pluralistic stories of their history or find ways of living together around shared concepts and connected subcommunities, then interaction* poses problems. For other disciplines that have grappled with similar issues, the stakes are threefold: (1) coherence, (2) status, and (3) progress (Beck & Stolterman, 2017). While it makes sense that mulling over matters of formal definition, or 'ironing things out', acts as a salve for these concerns (disciplining work), it is also possible to lose things when we do this. This is akin to proposing a set of guiding or 'big' questions for a discipline. Doing so may create the appearance of having achieved greater intellectual cohesion, yet, at the same time, it also thwarts creative efforts to take a field in new directions. If we agree on formal definitions, could this somehow be construed as self-imposing intellectual constraints and undermining the potential for more innovative or creative construals of interaction*?  We want to avoid such 'ironing work' and instead focus on examining the sea we swim in: language.

## 4. *For* interaction*: avoiding confusions and deflating problems

Although typically configured as matters of coherence, status, and progress, we argue that it is *not necessary* for HCI to see these things as problems. Instead, we want to make an argument for an orientation to interaction* that trades on its *language value*, as a promiscuous concept. This sidesteps a binary of 'interaction* definition' versus 'interaction* dispensing'. We believe that rethinking *the framing* of the perceived problems with interaction* as language troubles might help HCI communities' cohesion and channel further debates on HCI's chosen senses of interaction*, rather than another attempt at (re)articulating concepts. Here we do this in two ways. Firstly, we draw out a distinction between the ways interaction* is treated in formal research discourses and the ways it can be treated as a matter of practice—we do this to avoid confusion. Toward this end, we ask what *work* has interaction* done—and what work might it be currently doing—for HCI researchers?

Secondly, and in asking this, we attempt to deflate problems that have been argued dog interaction* by simply underlining how the same words can find distinct (and mostly untroublesome) meanings in different uses.

On the first point, the research we've discussed thus far tends to favour examining how interaction* plays out in HCI communities' academic, formal discourse: i.e., that found in the research record. H&O offer us a comprehensive and scholarly account, but this is only half the story. What we are missing here is research *practice*, i.e., the nitty gritty practical accomplishments involved in getting HCI research done. Maybe considering this can also tell us about interaction* as lubrication for research?

To put it another way, we are speaking of HCI research's necessarily local rationality and its mundane resolute, practical reasoning (see Lynch, 1993; Pollner, 1987). So, while H&O refer to a "vocabulary and [...] reasoning apparatus" of interaction, we want to consider how interaction* is embedded within the local reasonings of *this practice*: the situated, autochthonous 'languages of interaction*' or 'interaction* talk' that is major concern of the lifeworld of the institutions, labs, research groups, community networks, etc. that practice HCI. There are, of course, many different interactions* for different species of this HCI research practice and the places it is exercised in. There are thus ways of operationalising interaction* in community practice that do *not* presently reside in HCI's literature and cannot easily be subject to definition such as in H&O's schema. These then become ways of talking about interaction* into which practical reasoning about method and technologies become entwined—unstated—when they surface within the HCI community. Formal records of research tends to leave these practical reasonings 'on the cutting room floor'.

While one could imagine a programme of research that begins to articulate what those HCI practices are which embed interaction*, we have another more pertinent point to make from this. Contra H&O, we think interaction* has actually been *usefully* underdefined. This underdefinition is in support precisely of the everyday, dispersed and diverse treatments of interaction* in practice.

What does underdefinition look like? It is clear from H&O's instructive account that when HCI researchers *talk* about interaction* it can be used to say a great many things that may nevertheless in certain kinds of senses be 'incompatible' with one another. As we have stated before, this is generally not problematic; many interesting concepts cease to be interesting when precisely defined and doing so could render them less useful anyway (Schroeder, 2006, pp. 153-154). One might reasonably say—perhaps in a research paper if we confine ourselves to what is presently available to us—that someone tapping a touch screen is 'interacting'. Just as we can say that posting on social media is 'interaction'. Or that someone being physically tracked is 'interacting' with a geolocation system. Or we might say that 'interactions' are taking place with, around, or through technologies embedded in the social life of the home. Or, perhaps, we might even say someone using an urban public bicycle hire scheme is 'interacting' with networks not only of systems and data and other people but other kinds of discernible objects ("social, technical, scientific, intellectual, organizational, political, ethical worlds" (Taylor, 2015, p. 50)). Since such uses tend to imply intentionality we can also develop ways of talking about "unwitting" forms of interaction (Benford et al., 2006, p. 8; Reeves, 2011, p. 138) or perhaps "implicit interaction" (Ju et al., 2008). These different uses sit alongside one another and do not call out for policing or administrative conceptual work (cf. J&S, which presents some valuable ways of talking about a particular stance on interaction* yet also—very much unnecessarily in our view—argues for the need of disciplining "definitions and measures").

It seems unlikely to us that formulations like the above would cause a researcher reading a paper to squint and shake their head, as though using the word 'interact' (and its derivatives) to describe these different activities is in some way deeply incomprehensible to them; indeed, the unique characteristics of use might be "intuitively resonant" (Galle, 2011, p. 82) with readers as they bring their own senses of use to bear on what it is that others might be meaning. We do not deny that confusions *may* occasionally result. However, this is not ultimately resolvable through further definition (setting up intellectual straightjackets for future work) or through starting to abandon interaction* in favour of new paradigms (which may work against the value of retaining one of the few shared interests that features in HCI's complex cross-cutting communities). Further, as we have pointed out above in HCI practice, interaction* (gestalt: its language, its concepts) always appears in context. Even when left 'undefined' in formal research records, contextual details—such as the broader topic of a paper, the author(s), their home discipline(s), institutional affiliations, etc.—shape the way we might normally imbue interaction* with particular conventional meanings.

To reiterate, arguing *for* interaction* means accepting it as conceptually promiscuous: as a broker—embedded both (yet differently) in HCI research practices *and* its formal representation in the literature. Through this, interaction* helps manage the complex relationships between increasingly diverse research communities and influences that associate themselves in some way with HCI (primarily under the banner of conferences like CHI)[13].

## 5. A therapeutic reframing of interaction*

Interaction* in HCI is probably not a resolvable matter—and it need not be so. If instead we value interaction*'s capacity and permissiveness for absorbing different sensibilities, approaches, or perspectives, and if we focus *less* on agreements and more on embracing the possibility and generativity of *mis*understandings, then HCI will need to work out ways to surface, deal with, accommodate, and unpack those misunderstandings[14]. Firstly, we present a *therapeutic* attitude towards talking about troubles around interaction* and offer a sketch of how this might play out. Secondly, and more briefly, we reflect on how the structures of HCI might work against applying therapies to interaction*, leaving us with questions as to how HCI could be reordered to support such approaches.

The kind of therapeutic attitude we speak of turns on developing an appreciation for the role of language in use (in our case, within the HCI research community), and how the meaning of language, like 'the concepts of interaction', is formed *in and as* that use. This matter has been addressed previously by Wittgenstein, Ryle or Austin (amongst others; see Wittgenstein, 2009; Ryle, 1949; Austin, 1962). To begin, we can reread a portion of Wittgenstein's *Philosophical Investigations* (2009) as talking about how formalised versions of interaction* may play out in HCI's discourse ("a game"[15]) and what to do about it:

> Here the fundamental fact is that we lay down rules, a technique, for playing a game, and that then, when we follow the rules, things don't turn out as we had assumed. So we are, as it were, entangled in our own rules.

---

[13] We are hesitant to use Star and Griesemer's (1989) much over-used notion of a "boundary object" as it pertained (originally) to collaborating (scientific) communities centred around identifiable shared projects. We are cautious of stretching the notion (as others might have done) to describe the circumstances of broader community relations that we are speaking of.

[14] Consider calls for pluralistic design-oriented HCI (Bardzell, 2019), or, in comparison, territoriality over styles of ethnography exercised in HCI (Crabtree et al., 2009).

[15] Wittgenstein's sense of language games is often misinterpreted. We are not talking about a literal game (Schroeder, 2006, p. 129).

> This entanglement in our own rules is what we want to understand: that is, to survey. (Wittgenstein, 2009, §125)

Ryle and Austin's careful considerations of what it means to be "entangled in our own rules" in the course of language in use offers us some practical points on interaction*. Where Wittgenstein is therapeutic, Ryle adds a layer of the methodical (as does Austin), promoting an approach that takes words and their senses seriously by rolling them around enough to shake out both their meanings in use as well as their potential for (philosophical) conceptual confusions—with the latter to be avoided, mainly by reminding us of the former. Although primarily concentrating on conflicts between the philosophical and everyday modes in which 'the mental' and 'thinking words' feature[16], Ryle has a broader concern: "not how to apply [concepts], but how to classify them, or in what categories to put them" (Ryle, 1949, p. 49). Austin, in turn, considers how *doings* with words (e.g., promising, marrying, or betting) lead to a complex and potentially highly varied set of different *senses* in which things are said, and thus what is concretely being done, despite identical formulations of such statements. Statements of interaction* seem particularly vulnerable to the kinds of confusions outlined by Wittgenstein, Ryle and Austin, unless carefully unpacked in a particular way. This is *not* definitional work, though. Rather, it entails teasing apart vernacular from technical and distinguishing differences between the variety of senses interaction* takes, including our prior distinction between formal accounts in the literature and in routine research practices. *This* is what we mean when we say therapeutic.

Given the abstractness of the discussion so far, we want to now consider what a therapeutic approach could look like for HCI. What follows is a sketch.

There are many unspoken senses that articulations of interaction* seem to leverage. When we say technologies, systems, devices are 'interactive', we necessarily embed them within mundane social order—i.e., the accomplished stability of reality as the social world (Pollner, 1987). That is, we leverage *ordinary* understandings of interaction* to talk about what people do with and around computational technologies at the selfsame time as we (might) speak of what people do with other people. For instance, 'talking' is something we might say we are doing with people *or* speech-based technologies such as personal assistants, but we do not confuse one for the other[17] (Porcheron et al., 2018). The (always situated) *senses* of 'talking' do quite different things: talking in court (Atkinson and Drew, 1979) is quite different to talking with an intimate partner which is again quite different to talking to a machine (replete with ironic possibilities, cf. Reeves et al. (2019)). This blurring of interaction*— drawing of various concepts of interaction from our ordinary language—comes with a relevant repertoire of 'interaction words' like 'response', 'react', 'alert', 'remind', 'interrupt', 'share', 'detect' and so on.

---

[16] Arguing against Descartes (the "official doctrine"), Ryle pointedly illustrates that while philosophers generate "theory of the mind", 'anyone' can engage successfully in "describing the minds of others and in prescribing for them" (Ryle, 1949).

We talk about what interfaces, systems, devices do, or have done to them, in terms like this, but we have to keep in mind that these are also ways of talking about people individually, and jointly, using and living with machines. These ways of talking about machines are grounded in practical and everyday experiences of language in use, which means they borrow from the everyday sense but acquire technical senses in HCI discourses. As implied by H&O's catalogue of interaction*, different ways of talking about interaction* bring with them particular bodies of language, their associations, implications and insinuations. These different genres further imply more particular *families* of interaction words, also borrowing from an everyday sense.

Taking H&O's catalogue, consider "interaction as dialogue", which uses a language of turns, feedback, goals and sub-goals or "interaction as transmission" which speaks of noise, throughput, capacity, etc. "interaction as tool use" brings with it senses of what one might do with tools and toolic connections—tools break, tools amplify, etc. Considering "interaction as optimal behaviour", we enter the language of behaviour including costs and rewards, rational and irrational actions, and so forth. "Interaction as embodiment" speaks of intentions, context, situation, coupling, etc., while "interaction as experience" leverages a host of normative reasonings about 'feeling words', such as: emotions, surprise, stimulation, and also aesthetics, expectation, etc.

In a therapeutic approach, HCI's language of interaction* acts as a *provocation* to facilitate the work of its conceptual promiscuity, which includes: foregrounding and supporting the exploration of conceptual confusions, surfacing the troubles of categorisation, reminding us of the play of metaphor, and foregrounding the prevalent leverage of mundane senses of terms, which might then be played with 'fast and loose'. By engaging in this work, we can promote more generative discussions and greater depth of understanding of interaction* in HCI. Returning to our family of affiliated 'interaction words' ('response', 'react', etc.) let us consider 'interrupt', a key interaction mechanism in HCI (Janssen et al., 2015). What does 'interrupt' mean specifically and what might the different senses be for an app, an agent, a personal device, an Internet of Things / device 'ecosystem', a social network recommendation, a friend's message, or an organisation's call-to-action to 'interrupt' us?

---

[17] Except, perhaps in some exceptional circumstances we could concoct.

There are many senses in which we might make 'interrupt' meaningful, and it's likely that you, the reader, are constructing scenarios to make sense of 'interrupt' yourself. Does it make sense to categorise a device and an IoT ecosystem as interrupting us in the same way? What about push notifications? Do they interrupt in the same way as a productivity app reminding us to 'get back to work'? We can say a social network interrupts us to deliver a recommendation of some kind, but is this sense of 'interruption' the same as that produced by a message from a friend? When we say an agent interrupts someone, do we mean it in the same sense as an interruption on the street by a police officer? Something could be categorised as an interruption by an app's design(er) but treated as a welcome intervention, for example, by the 'receiver' of that interruption. Overall, we might wonder what it means for us to use interaction* as a frame within which to talk about 'interruptions as interactions' that are generated or mediated by different kinds of technologies in different circumstances—things which ordinarily take much of their sense from mundane, everyday sociality. When we examine the literature on interruption, it is often unclear what corresponding sense of interruption might be meant. For instance, we might point to the Special Issue edited by Janssen et al. (2015), in which something called "self-interruption" is subsumed into a general sense of interruption, a concept itself that is also left undifferentiated or textured.

Our point here is that interaction*—its families of languages and concepts—are *far* more complex than any of the interaction* debate has presently articulated. The therapeutic approach is a way to *live with* misunderstandings, confusions, and conflations, while resisting urges to work toward resolving epistemic differences.

To close, we want to raise HCI's structural, disciplinary conundrums and the interaction* debate. Firstly, we are reluctant to recommend any distinct ways 'forward'. The idea of structural progress is embedded in how we talk about interaction* in HCI. If we are proposing an embrace of interaction*'s promiscuity, this might also mean potentially doing away with—or at least being cognisant and wary of—various normative discourses of scientific progress (i.e., replicability, cumulation, forward motion, etc.; Reeves (2015)). Instead, we restrict ourselves to observations and open questions about HCI's organisation, which we feel are more generative than the provision of implications.

The present structural organisation of HCI (as with many academic communities) tends towards bifurcation. Premier examples of this include the ACM CHI conference's subcommittee structure[18] which represents a particular solution to the practical problem of scaling participation rates that we mentioned earlier. However, this kind of structural bifurcation also mixes domain, method, epistemology, and contribution. Structures like this might invite rigidity or confusion in HCI, offering little provision for managing differences in interaction*. We then might ask: how could we rethink such structures to foreground and make differences in interaction* productive? Present structures seem more likely to promote conflict and confusion.

## 6. Conclusions

A multifaceted discourse composed of interaction*—its attendant concepts and languages—prevails as a concern in HCI. Rarely is this explicit, and mostly it is implied. As sets of concepts, interaction* is about research practices as much as it is about the words HCI researchers use when they generate talk and texts. Interaction* acts as a common recognisable interest, and it is perhaps the only thing identifiably bridging gaps between increasingly disparate HCI communities. Perhaps HCI might be better off being more cognisant of the multiplicity of ways in which interaction*—admittedly confusing at times—lets us talk about what is a massively varied focal interest. It is probably worth pausing for thought before we either seek to tighten its useful lack of definition or dispense with it in view of new conceptual frontiers. We think there might be value in rediscovering interaction* for its sense of intellectual promiscuity. We believe it is perfectly reasonable to hold multiple, potentially contradictory senses of interaction*, get along with each other, *and* be critically productive as HCI-affiliated communities. But we probably need something like therapies to help us *live with* our frictions rather than finding ways to eliminate them through strategies of discipline. However, present structures of HCI might hinder this, so we perhaps should begin asking the question: how might we reorganise our HCI communities to facilitate interaction*?

---

[18] As of CHI 2019, these were as follows: User Experience and Usability, Specific Application Areas, Learning, Education and Families, Interaction Beyond the Individual, Games and Play, Privacy, Security, and Visualization, Health Accessibility and Aging, Design, Interaction techniques, Devices and Modalities, Understanding People: Theory, Concepts, Methods, Engineering Interactive Systems and Technologies.

## Acknowledgements

This work is supported by the Engineering and Physical Sciences Research Council [grant numbers EP/M02315X/1, EP/K025848/1]. We gratefully acknowledge valuable feedback and input from Antti Oulasvirta, Alex Taylor, and Jocelyn Spence when writing this paper, as well as insightful comment from anonymous reviewers.

Ludwig Wittgenstein. 2009. *Philosophical investigations*, 4th edition, P. M. S. Hacker and Joachim Schulte (eds. and trans.), Oxford, UK: Blackwell.